\documentclass[conference]{IEEEtran}

\ifCLASSINFOpdf
 
\else
 
\fi

\usepackage[noadjust]{cite}

\usepackage{amsmath,amssymb}
\usepackage{graphicx}
\usepackage{braket}
\usepackage{epsfig}
\usepackage{epstopdf}
\usepackage{mdwmath}
\usepackage{xcolor}
\usepackage{varwidth}
\usepackage{framed}
\usepackage{float}
\usepackage{placeins}
\usepackage{afterpage}
\usepackage{ragged2e}
\usepackage{amssymb}
\usepackage{pifont}

\usepackage{blindtext}
\usepackage{enumitem}
\usepackage{xcolor}
\usepackage{eqparbox}
\usepackage{fixltx2e}
\usepackage{multirow}
\usepackage{enumitem,lipsum}

\usepackage[english]{babel}
\newcolumntype{P}[1]{>{\centering\arraybackslash}p{#1}}
\usepackage{graphicx}
\usepackage{amsmath, amssymb}
\usepackage{graphicx}
\usepackage{epsfig}
\usepackage{epstopdf}
\usepackage{array}
\usepackage{graphicx}
\usepackage{subcaption}
\usepackage{hyphenat}

\usepackage{bbm}

\usepackage{booktabs} 
\usepackage{graphicx}
\usepackage{caption}
\usepackage{bbding}
\usepackage{pifont}
\usepackage{wasysym}
\usepackage{amssymb}

\usepackage{xcolor}         
\definecolor{blue}{rgb}{0.06, 0.2, 0.65}

\usepackage{framed,url,caption,graphicx}
\usepackage{soul,xspace,tabularx}

\expandafter\def\expandafter\UrlBreaks\expandafter{\UrlBreaks
  \do\a\do\b\do\c\do\d\do\e\do\f\do\g\do\h\do\i\do\j%
  \do\k\do\l\do\m\do\n\do\o\do\p\do\q\do\r\do\s\do\t%
  \do\u\do\v\do\w\do\x\do\y\do\z\do\A\do\B\do\C\do\D%
  \do\E\do\F\do\G\do\H\do\I\do\J\do\K\do\L\do\M\do\N%
  \do\O\do\P\do\Q\do\R\do\S\do\T\do\U\do\V\do\W\do\X%
  \do\Y\do\Z}

\newcommand{\eat}[1]{}

\newcommand{\edprotocolssimulator}{~\url{https://public.after.acceptance/ed_protocols_simulator}~}

\usepackage{braket}
\usepackage[utf8]{inputenc}
\usepackage{algorithm2e}
\makeatletter

\begin{document}


\title{Analysis of Asynchronous Protocols for Entanglement Distribution in Quantum Networks}

\author{\IEEEauthorblockN{Shahrooz Pouryousef}
\IEEEauthorblockA{ 
 Cisco Research \& UMass Amherst\\
USA\\
 shahrooz@cs.umass.edu}
\and
\IEEEauthorblockN{Hassan Shapourian}
\IEEEauthorblockA{Cisco Research\\ USA\\
hshapour@cisco.com}
\and
\IEEEauthorblockN{Don Towsley}
\IEEEauthorblockA{UMass Amherst\\ USA\\
towsley@cs.umass.com}
}

\maketitle

\IEEEpeerreviewmaketitle

\begin{abstract}
The distribution of entanglement in quantum networks is typically approached under idealized assumptions such as perfect synchronization and centralized control, while classical communication is often neglected. However, these assumptions prove impractical in large-scale networks. In this paper, we present a pragmatic perspective by exploring two minimal asynchronous protocols: a parallel scheme generating entanglement independently at the link level, and a sequential scheme extending entanglement iteratively from one party to the other. Our analysis incorporates non-uniform repeater spacings and classical communications and accounts for quantum memory decoherence. We evaluate network performance using metrics such as entanglement bit rate, end-to-end fidelity, and secret key rate for entanglement-based quantum key distribution. Our findings suggest the sequential scheme's superiority due to comparable performance with the parallel scheme, coupled with simpler implementation. Additionally, we impose a cutoff strategy to improve performance by discarding attempts with prolonged memory idle time, effectively eliminating low-quality entanglement links. Finally, we apply our methods to the real-world topology of SURFnet and report the performance  as a function of memory coherence time.
\end{abstract}

\section{Introduction}

Quantum networks represent a pivotal technology for a multitude of applications, including quantum-safe secure communication~\cite{bennett2020quantum,peev2009secoqc,wang2014field,stucki2011long,giovannetti2004quantum}, distributed quantum computing~\cite{cirac1999distributed}, and several others~\cite{kimble2008quantum,wehner2018quantum}. The scalability of these networks faces a significant obstacle due to photon loss. In response, entanglement distribution networks~\cite{briegel1998quantum,munro2015inside,azuma2022quantum} have emerged as a solution to mitigate this challenge. 

The proof of concept experiments on the entanglement distribution across a small network up to three nodes has been achieved~\cite{jing2019entanglement,yu2020entanglement,pompili2021realization,hermans2022qubit,van2022entangling,knaut2023entanglement}, but there is still a long way to reach large scale networks.
As entanglement distribution networks scale up, 
establishing and maintaining reliable and high-fidelity entanglement across the network
require developing novel protocols~\cite{lloyd2004infrastructure,dahlberg2019link} for routing while dealing with noisy quantum hardware.

Entanglement distribution is often envisaged as a two-way generation process where a pair of users demand a shared entanglement between them. 
As a result, a common approach involves generating end-to-end entanglement by in advance reserving the necessary resources on a given path for a pair of users and trying all links in perfect harmony. These approaches are collectively known as synchronous or time-slotted~\cite{pant2019routing,li2021effective,kozlowski2020designing} and often rely on a central controller for network orchestration~\cite{skrzypczyk2021architecture}.
However, synchronous approaches are challenging to implement at large scale (e.g., networks of networks) due to increasing workload on the central controller and the need for perfect synchronization. 

\begin{figure}
\centering
    \includegraphics[width=\columnwidth]
    {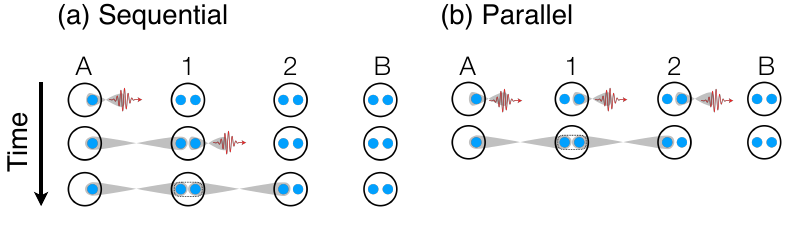}
    \vspace{-0.2in}
  \caption{Two asynchronous protocols for entanglement distribution studied in this paper.} \label{fig:ed_protocols}
\end{figure}

Given that, asynchronous approaches~\cite{kamin2023exact} are proposed as a scalable solution with the possibility of distributed routing~\cite{yang2024asynchronous,xiao2023connectionless,chen2023q,wang2022pre}. In this paper, we study two types of asynchronous approaches: Sequential and parallel as shown in Figure~\ref{fig:ed_protocols}. The main difference between these protocols and synchronous ones is that elementary entanglement generation is attempted independently for each direct link; hence, there is no need for perfect arrangement so that qubits arrive at each repeater precisely at the same time or within a short time window.
Unlike previous studies, we not only include the effect of noisy quantum memory  in our analysis but also take into account non-uniform repeater spacing and latencies due to classical communication. To avoid further complexity however, we focus on a \emph{minimal} hardware setup based on first-generation repeaters, where there is only one memory (within each repeater) per optical link connecting two quantum repeaters and the entanglement sources are located within each repeater. 

Lastly, we make some realistic assumptions about how the link-level entanglement is generated: 
Our protocols differ from previous ones in 
that each user pair consists of a sender and a receiver, where the sender is the party that initiates the establishment of an EPR pair with the other party, the receiver. 
We further consider each repeater to be equipped with an entangled photon source and detector that can establish a link-level entanglement with its neighboring nodes (consistent with our first assumption of directed paths from sender to receiver). This is in contrast with common schemes where an entangled photon source or detector is placed in the middle between every pair of repeaters, which may have a better generation rate compared to our setup albeit at the cost of introducing extra overhead in terms of hardware and quantum signal processing to ensure photon phase matching and possibly more rounds of classical communication.
As we show, despite this apparent asymmetry the overall performance remains symmetric (see Figure~\ref{fig:one_repeater} as an example and Section~\ref{sec:One repeater} for details).

We develop two simulation methods—discrete event simulator and Monte Carlo method—to analyze the protocols in different scenarios 
including performance as a function of repeater locations in a one-repeater chain, analysis of optimal secret key rate as a function of distance and coherence time for fixed number of uniformly distributed repeaters, and comparison of the performance of each protocol as a function of coherence time on a random linear network and a real-world network (SURFnet). 
We verify that the results of the two simulators match. Overall, we find that although the parallel protocol outperforms the sequential protocol, the difference is not so sufficient to overcome the challenges of implementing parallel protocol compared to the sequential one. This suggests the potential of sequential protocols as primary candidates for realizing the quantum Internet.

The rest of our paper is organized as follows: In Section~\ref{sec:protocols}, we start by explaining the network model, the noise model and general assumptions, and next we present the two asynchronous protocols and our cutoff strategy. 
Section~\ref{sec:evaluation} is devoted to various evaluations as mentioned above.
We discuss the previous related work in Section~\ref{sec:related work}
and close with some concluding remarks and future research directions in
Section~\ref{sec:conclusion}.




\section{Entanglement distribution protocols}
\label{sec:protocols}

In this section, we explain our physical setting and noise model. We discuss how we include classical communication delay and cutoffs for quantum memories for entanglement distribution in sequential and parallel protocols.


\subsection{Network model}

We represent a quantum network  by a graph $G = (V, E)$, where $V$ is the set of nodes and $E$ is the set of optical fiber links. Between each pair of nodes, there is a path (a sequence of repeaters) that can be considered as a repeater chain. We assume each repeater has only one quantum memory per interface. In this paper, we assume paths between user pairs are disjoint and the problem of congestion control in quantum networks \cite{chen2023q} is out of the scope of this paper. 


When a repeater attempts to generate entanglement with another one, it locally prepares a two-qubit Bell pair (or Einstein-Podolsky-Rosen (EPR) pair), stores one qubit in the quantum memory and sends the other qubit in the form of a photon to the target repeater. Upon receiving the photon, the target repeater stores the qubit in its local quantum memory and sends an acknowledgement to the sender. This process is called elementary link (or link-level) entanglement generation. Once two quantum memories within a repeater, which are  assigned to a path, are entangled with the two neighboring repeaters, a Bell state measurement (or entanglement swapping) can be performed to generate a longer entanglement link connecting the two neighboring repeaters. We will explain our protocols in more detail as to how  link-level entanglement generation and swaps are scheduled.

\subsection{Noise model and assumptions}
\label{sec:noise model}

Due to photon loss in the fiber,  link-level entanglement generation is probabilistic, and the success probability is given by
\begin{align}
    \label{eq:link-level-prob}
    p_i = p_\text{link} e^{-\alpha L_i},  
\end{align}
where $\alpha$ is the fiber attenuation rate, $L_i$ is the length of the fiber for the $i$-th link, and $0\leq p_\text{link}\leq 1$ accounts for fiber coupling efficiency, detector efficiency, etc. Here, we set $\alpha= 0.046~[\text{km}^{-1}]$ corresponding to $0.2$~dB/km signal attenuation in fiber and $p_\text{link}=1$ for simplicity although we experiment the affect of different values of $p_{link}$ on overall entanglement generation rate in section ~$\S$\ref{sec:evaluation}.
We take the speed of light in the fiber to be $c=2\times 10^8$~m/s to calculate potential delays associated with quantum signal transmission and classical communications through different entanglement protocols.

We formulate the entanglement generation process across $n$ repeaters in terms of $n+1$ random variables $N_i$ associated with $n+1$ links. $N_i$ denotes the number of attempts until a successful generation of an EPR pair for the $i$-th link connecting $(i-1)$-th and $i$-th repeater nodes and follows the geometric distribution $p_i (1-p_i)^{N_i-1}$ with $N_i > 0$.

We use error models for the quantum memory, noisy generation of link-level entanglement, and noisy Bell-state measurements introduced in~\cite{kamin2023exact}. We briefly review them here.

A noisy local EPR pair (described by density matrix $\hat \rho_0$) is characterized by the fidelity $F$ and given by
\begin{align}
\label{eq:link-level-epr}
    \hat \rho_0 = F |\Psi_+\rangle\!\langle \Psi_+| + (1-F) |\Psi_-\rangle\!\langle \Psi_-|,
\end{align}
where $\ket{\Psi_\pm} = (\ket{10}\pm\ket{01})/\sqrt{2}$, and $F=1$ represents a perfect state generation, i.e., the state $\ket{\Psi_+}$ is used as the ideal Bell pair. 
We also consider a two-qubit depolarizing channel
for elementary link generation and Bell-state measurement at the repeaters as in
\begin{align}
    \label{eq:2q-depolarizing}
    \tilde \Gamma_\mu (\hat \rho) = \mu \hat \rho 
    + (1-\mu) \frac{\hat {\mathbbm{1}}}{4},
\end{align}
which is characterized by the parameter $0\leq \mu \leq 1$. Here, we use  density matrix $\hat \rho$ to denote the state of the system.
To account for noisy quantum memories, we consider a dephasing channel such that when a quantum memory stores a qubit $a$, and remains idle for time $t$, it undergoes a dephasing channel
\begin{align}
    \label{eq:memory-dephasing}
    \Gamma_t (\hat \rho) = \left(\frac{1+e^{-t/\tau_\text{coh}}}{2} \right) \hat \rho 
    + \left(\frac{1-e^{-t/\tau_\text{coh}}}{2} \right) \hat Z_a \hat \rho \hat Z_a,
\end{align}
where $\tau_\text{coh}$ denotes the memory coherence time, and $\hat Z_a$ is the $Z$ Pauli operator acting on qubit $a$.

We now derive the full noise model for an end-to-end entanglement generation. The starting point is the elementary link entanglement generation (preparation of a local EPR pairs and transmission of one qubit over the channel) which is given by
\begin{align}
    \hat \rho_i = \tilde\Gamma_{\mu_i} (F_i \ket{\Psi_+}\!\bra{\Psi_+}+ (1-F_i) \ket{\Psi_-}\!\bra{\Psi_-} ).
\end{align}
Here, we use the subscript $i$ to denote the $i$-th link.
In each iteration, after swapping in $n$ repeaters, we obtain an end-to-end EPR pair given by
\begin{align}
    \label{eq:final-state}
    \hat \rho_{e2e} = \tilde\Gamma_{\mu_{e2e}} \left[ f_{e2e} \ket{\Psi_+}\!\bra{\Psi_+}+ (1-f_{e2e}) \ket{\Psi_-}\!\bra{\Psi_-} \right],
\end{align}
where
\begin{align}
    \label{eq:mu-e2e}
    \mu_{e2e} =& \mu^{n} \prod_{i=1}^{n+1} \mu_i, \\
    f_{e2e} =& \frac{1}{2} + \frac{1}{2} 
    \prod_{i=1}^{n+1} (2F_i-1) \cdot \nonumber \\
    \label{eq:small-f-e2e}
    &\exp\left[-\frac{1}{\tau_\text{coh}}\left(T_A + T_B + \sum_{i=1}^{n} (t_{i,L}+t_{i,R})\right)\right],
\end{align}
$\mu_i$ and $\mu$ denote the noise parameter in (\ref{eq:2q-depolarizing}) associated with the $i$-th link entanglement generation and entanglement swapping at $i$-th repeater, respectively.
For the memory idle times $t_{i,s}$, we use two subscripts: $i$ to denote the two quantum memories in $i$-th repeater and $s=L,R$ to indicate the left and right quantum memories on a given path (see Figure~\ref{fig:ed_protocols}). $T_A$ and $T_B$ denote the idle time of quantum memories in the sender and receiver, respectively.



\subsection{Secret key rate and Fidelity}

The performance of entanglement distribution network is measured by two quantities: First, the average end-to-end entanglement generation rate (or ebit rate), $R_{e2e}$, i.e., how often a successful EPR pair is established between the two users.
Second, the average end-to-end fidelity, $F_{e2e}$, 
which is a proxy for the quality of the end-to-end EPR pairs and is defined as the final state fidelity,
\begin{align}
    \label{eq:fidelity}
    F_{e2e} = \overline{\bra{\Psi_+}\hat \rho_{e2e}\ket{\Psi_+}}= \mu_{e2e} \overline{f_{e2e}} + (1-\mu_{e2e} )/4,
\end{align}
where $\hat \rho_{e2e}$ defined in (\ref{eq:final-state}) and symbols with bars denote the mean value averaged over many iterations. 

Using the end-to-end generation rate and fidelity, various quantum network utility functions can be constructed~\cite{vardoyan2022quantum}.
For a security application of quantum networks, we consider entanglement-based quantum key distribution (QKD) (also known as E91 protocol~\cite{ekert1991quantum}). The secret key rate (SKR) is given by
$S = R_{e2e}\cdot r$, where $r$ is the secret fraction,
\begin{align}
    \label{eq:secret-fraction}
    r = 1 - h(e_x) - h(e_z),
\end{align}
where
$h(p) =  -p \log_2 p - (1-p) \log_2 (1-p)$
is the binary entropy function, and
the average qubit error rates (QBER) in Z and X bases are given by
\begin{align}
    \label{eq:qber}
    e_z &= (1-\mu_{e2e} )/2, \nonumber \\
    e_x &= (1 + \mu_{e2e} )/2 -  \mu_{e2e}  \overline{f_{e2e}}.    
\end{align}
We note that in QKD the users do not need to wait until the end-to-end entanglement is established and can measure their respective qubits as soon as they have them. As a result, we set the end-user memory idle times $T_A=T_B=0$ in (\ref{eq:small-f-e2e}).

In the rest of this section, we present the two entanglement distribution protocols in more detail and discuss a cutoff strategy to prevent successful events with long duration resulting in generation of end-to-end Bell pairs with low fidelity.

\subsection{Sequential protocol}

The sequential entanglement generation protocol starts with a user (sender) demanding to establish an EPR pair with another user (receiver). As such, the sender prepares a local EPR pair and transmits one of the qubits to the first repeater. The first repeater sends a (classical) acknowledgement signal to the sender while generating a local EPR pair and sending it to the second repeater. 
The second repeater repeats this process upon receiving a qubit from the first repeater. As soon as the first repeater receives the acknowledgement signal from the second repeater, it performs a Bell-state measurement to generate a direct EPR pair between the sender and the second repeater. The measurement outcomes are then transmitted to the sender.
This process continues until we arrive at the receiver, and hence we refer to this as the sequential protocol. Figure~\ref{fig:ed_protocols}(a) shows the first two steps for a  linear (chain) network with two intermediate repeaters until we obtain an EPR pair between the sender and the second repeater.
The duration of each attempt in this protocol is given by 
\begin{align}
    \label{eq:T-sequential}
    T = \sum_{i=1}^{n+1} 2\tau_i N_i,
\end{align}
where $\tau_i=L_i/c$ is the signal travel time over the $i$-th link.
As a result, the average ebit rate
is found to be $R_{e2e}=1/\overline{T}$ where
\begin{align}
    \label{eq:seq-mean-time}
    \overline{T} &= \sum_{i=1}^{n+1} \frac{2\tau_i}{p_i},
\end{align}
we use $\overline{N_i}=1/p_i$ (c.f.~Section~\ref{sec:noise model}), and $p_i$ denotes the link-level entanglement generation success probability (\ref{eq:link-level-prob}). More details of how we calculate the ebit and secret key rates are provided in Appendix A.

We note that the sequential protocol is easily adapted to handle an arbitrary network topology and can be implemented as a hop-by-hop process with a distributed routing as proposed in \cite{chen2023q,li2022connection}. Therefore, it should be considered as a plausible candidate for 
connectionless entanglement distribution in quantum networks which is analogous to the best-effort methodology of today's Internet.

\subsection{Parallel protocol}

Similarly, the parallel entanglement generation protocol starts by a user (sender) demanding to establish an EPR pair with another one (receiver). In this case, all repeaters on a directed path connecting the sender to the receiver simultaneously attempt to generate link-level entanglements with their successive repeater (see Figure~\ref{fig:ed_protocols}(b)), and hence the namesake parallel. Each repeater then sends a classical acknowledgement signal to the corresponding repeater from which they receive a qubit. Once both quantum memories within a repeater is successfully entangled with two other repeaters, the Bell-state measurement is performed, and the outcome is passed on to the sender. An iteration is finished when the sender receives all Bell-state measurement outcomes. Therefore, the duration of each attempt is found by
\begin{align}
    \label{eq:T-parallel}
    T = \max (\{T_i + \sum_{j\leq i} \tau_j | 1\leq i \leq n \}),
\end{align}
where  $T_i= \max((2N_i-1) \tau_i, 2 N_{i+1} \tau_{i+1})$ is the time when the swap is carried out at the $i$-th repeater. The apparent asymmetry in the max function arguments is due to the fact that quantum signals arrive from left to right as shown in Figure~\ref{fig:ed_protocols}(b).  We use Monte Carlo method to numerically compute the average duration $\overline{T}$ and the ebit rate $R_{e2e}$ as deriving a closed-form expression is cumbersome beyond the one repeater case.
Explicit expressions for the memory idle times and other details of calculations are provided in Appendix B.


A potential distributed implementation of the proposed parallel protocol involves additional information exchange with neighboring nodes, as suggested by \cite{yang2024asynchronous}. Alternatively, another approach could entail utilizing one of the multiplexing schemes presented by \cite{aparicio2011multiplexing} and applying the parallel scheme independently for each user pair during specific time slots. In both scenarios, if repeaters adopt a cutoff strategy (as detailed in the subsequent section), there might be a need for some classical messaging between nodes in the event of qubit expiration, indicating the initiation of a new round of the protocol.

We envision that future quantum networks may incorporate both sequential and parallel protocols but at different layers of the network stack \cite{dahlberg2019link,kozlowski2020designing}. Specifically, it seems plausible that physical links would continuously attempt to generate link-level entanglements at the network's physical layer. These link-level entanglements would then be utilized by a network layer protocol implementing a sequential scheme to establish end-to-end entanglement. Consequently, the network layer, utilizing a sequential protocol, would not have to wait for the generation of a link-level entanglement, akin to how the link layer in classical Internet is responsible for transferring packets to the next hop without engaging in routing tasks.

\subsection{Cutoff strategy}



The basic versions of the above protocols involve attempting link-level entanglement generation until successful.
This can yield a high ebit rate but with end-to-end Bell pairs having low fidelity.
In any case, we work with the average ebit rate which remains finite. 
Thus, it makes sense to introduce a cutoff time $\tau_\text{cut}$ for quantum memories beyond which we halt the generation process and drop the Bell pair. 

We note that we apply this cutoff to each memory separately as opposed to overall cutoff including all memories. Our reasoning is that the latter is not practical as the total idle time over all memories is global information and precludes any possibility of a distributed protocol.
We still keep track of memory idle time for the successful events and calculate the final state according to (\ref{eq:final-state}). This is similar to the treatment in~\cite{van2020extending,rozpkedek2018parameter} while
 it differs from
the hard decoherence time approximation~\cite{collins2007multiplexed,shchukin2019waiting,pouryousef2023quantum} where quantum memory is assumed to remain perfect until a critical time after which it is suddenly destroyed.
The cutoff strategy clearly sacrifices the ebit rate; however, the successfully generated Bell pairs have larger end-to-end fidelities. We show the derivation details of the sequential protocol with cutoff in Appendix A and explain how we numerically evaluate the performance of the parallel protocol with cutoff in Appendix B.




\section{Evaluation}
\label{sec:evaluation}

In this section, we evaluate the effectiveness of the sequential and parallel protocols designed for repeater chains. Initially, we present the performance results of the protocols applied to a single repeater chain. Subsequently, we analyze the optimal secret key rate derived from the optimal cutoff value for repeaters in these two protocols. Following that, we examine the behavior of the two protocols when $n$ repeaters are randomly positioned along the repeater chain. Finally, we evaluate the protocols in the context of a real network (SURFnet obtained from \cite{knight2011internet}). 

To numerically evaluate the network performance in terms of ebit rate and SKR, we have developed two independent codes: a Monte Carlo simulation to evaluate the average quantities
based on the equations containing random variables (e.g., (\ref{eq:T-sequential}) and (\ref{eq:T-parallel})) and a discrete event simulator to sanity-check the equations and results in this paper. The simulator is open-source and available at
\edprotocolssimulator.


For most of the simulations, we set the $\mu_i=\mu = 1$ and $F_i=1$ in (\ref{eq:mu-e2e}) and (\ref{eq:small-f-e2e}) for simplicity, and focus exclusively on the impact of coherence time $\tau_\text{coh}$ and cutoff time $\tau_\text{cut}$. Furthermore, we explore the effect of $\mu_i=\mu$, $F_i$, and $p_\text{link}$ on the feasible regime for QKD in two different protocols.

\subsection{One repeater inhomogeneous repeater chain}
\label{sec:One repeater}

\begin{figure}
\centering
    \includegraphics[width=9cm]{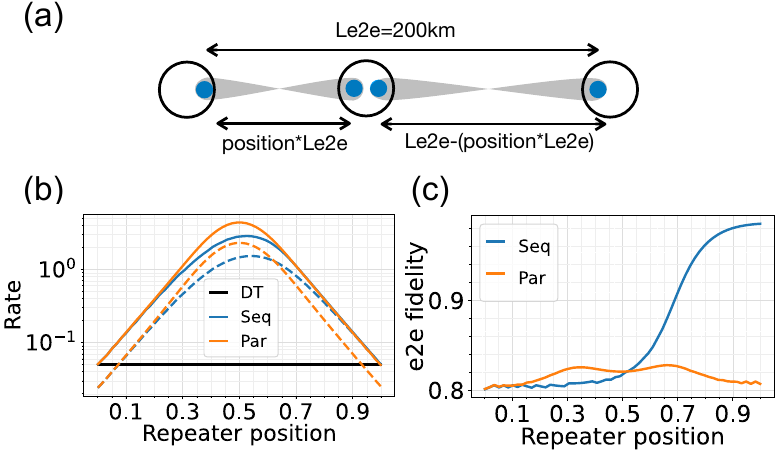}
    \vspace{-0.06in}
  \caption{A repeater chain with one repeater. Solid lines show the ebit rate and dashed lines show the SKR of the associated protocols. For each color, the solid line indicates the ebit rate and the dashed line represents the SKR.} 
  \label{fig:one_repeater}
\end{figure}

In this experiment, we examine a single repeater chain, as illustrated in Figure~\ref{fig:one_repeater}(a). We refer to the link between the source and the repeater as the left link and the other one as the right link. The repeater's position is varied between the end nodes, and we plot the ebit and secret key rates for the sequential, parallel, and direct transmission schemes. The end-to-end distance $L_{e2e}$ is set at $200$ kilometers, while coherence and cutoff times are fixed at $0.1$ and $0.05$ seconds, respectively.

Several remarks about the results in Figure~\ref{fig:one_repeater} are in order. Figure~\ref{fig:one_repeater}(b) indicates that the parallel protocol is not always better than the sequential protocol with respect to the SKR. Specifically when the repeater is close to the receiver, the sequential protocol outperforms the parallel protocol. When the repeater is positioned too close to either end node, both protocols yield similar ebit rates. The reason is in this regime the ebit rate is mostly limited by the longer link and the setup is effectively repeaterless although the performance still remains better than the direct transmission since the longer link is still shorter than the end-to-end distance. The gap between the ebit rate for the two protocols increases as the repeater position approaches the center of the link. Unlike the case where the repeater is in the middle of the line, the symmetric point for the sequential scheme differs. 
On the other hand, in Figure~\ref{fig:one_repeater}(c), we note an improvement in the end-to-end fidelity of delivered EPR pairs when employing the sequential entanglement generation protocol and moving the repeater closer to the receiver node, while the end-to-end fidelity for the parallel scheme remains constant. This is attributed to the sequential protocol, where the right link refrains from attempting to establish a link-level entanglement with the end node until it receives the photon from the source node. In case of successful photon reception, the entanglement attempt is made. In contrast, the parallel scheme generates entanglement on the right link immediately due to its short length. However, the left link requires multiple attempts to generate a link-level entanglement, causing qubits at the right memory of the repeater to decohere and decrease fidelity. This issue is absent in the sequential scheme, as entanglement on the right link is generated only when the left link successfully establishes entanglement. Another consequence of this phenomenon appears in Figure~\ref{fig:one_repeater}(b) where the secret key rate approaches the ebit rate as the repeater is moved towards the right end node since the secret key fraction is nearly one (or nearly vanishing error rates). 


\subsection{Cutoff vs coherence time vs distance}

In this experiment, we consider $7$ repeaters deployed on a link between a source and an end node. The goal of this experiment is to show how much gain we obtain from improving the cutoff or the coherence time and how the SKR can change if the repeaters are placed on a link with different lengths.

\begin{figure}
\centering
    \includegraphics[width=\columnwidth]{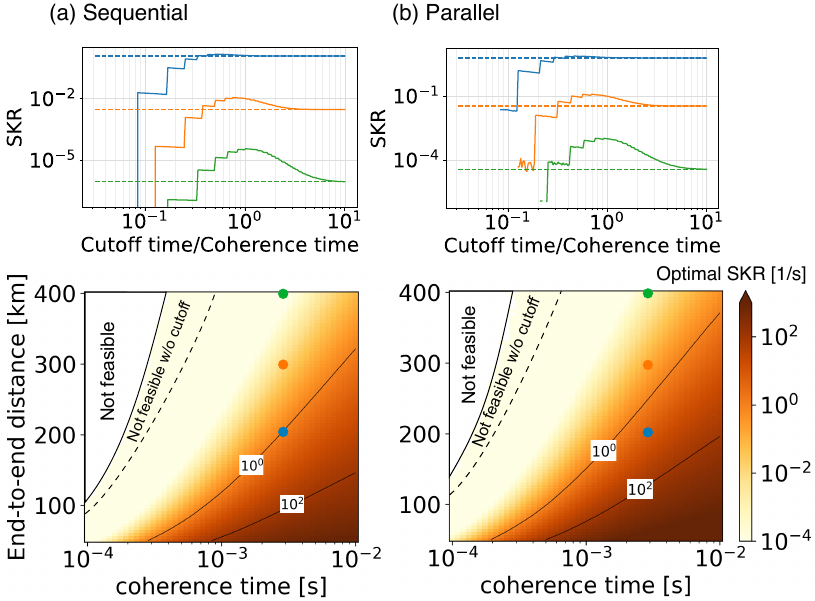}
    \vspace{-0.06in}
  \caption{The upper panels show sample plots of the SKR as a function of cutoff using $7$ repeaters on three different distances (points in the lower panels (color coded)) for $L_{e2e}=200,300$ and $400$ km (blue, orange, green), respectively. The dashed lines in the top plot indicate the SKR without cutoff for the corresponding distance indicated in the bottom plots. The lower panels show maximal SKR for a given distance and coherence time by optimizing the cutoff. We show two contour lines at $10^0$ and $10^2$ SKRs as a guide to eyes. The infeasible regions correspond to the case error rates are so large that the secret fraction (\ref{eq:secret-fraction}) vanishes.} 
  \label{fig:cutoff_coherence_SKR}
\end{figure}

The upper panels in Figure~\ref{fig:cutoff_coherence_SKR} show sample plots of the SKR as a function of cutoff using $7$ repeaters on three different distances (points in the lower panels (color coded)) for $L_{e2e}=200,300$ and $400$ km (blue, orange, green), respectively for the sequential and parallel protocols with a fixed coherence time of $0.003$ seconds. The cutoff value is progressively increased. It is evident that augmenting the cutoff value can be advantageous for the SKR, but there is a threshold beyond which it may adversely impact the SKR by leading to the delivery of EPR pairs with lower fidelity. A shorter cutoff, on the other hand, can mitigate this issue. The step-like behavior of the SKR in both schemes is associated with the one round of communication required for entanglement generation on each link. In essence, if the cutoff value allows for a sufficiently long duration for one round of communication on a link, it may enhance the rate and, consequently, the SKR. However, merely increasing the cutoff may not be adequate for achieving this improvement in one round of communication.

The lower panels in Figure~\ref{fig:cutoff_coherence_SKR} display a color plot representing the optimal SKR as a function of distance and coherence time in a repeater chain for the sequential and parallel protocols, respectively. We obtain the optimal SKR at each point in this plot by calculating a similar plot as in the upper plots and finding the cutoff value which yields the maximum SKR. Indeed, upper panels show examples of these curves for the three points (color coded) marked in lower plots. Both sequential and parallel protocols exhibit three distinct regions. The first region denotes circumstances where it is impossible to achieve any secret key rate due to a combination of low coherence time and long distance (and hence long delay due to classical communication) between end nodes. Notably, this region is smaller for the parallel protocol. As coherence time increases, a second region emerges where a non-zero secret key rate is possible, but it requires the application of cutoffs. The third region is where memory coherence time is high enough to deliver non-zero SKR possibly even without using cutoffs. 

\subsection{Feasible region for imperfect repeaters}

As shown in the lower panels of Figure~\ref{fig:cutoff_coherence_SKR}, the SKR degrades as we go to shorter memory coherence time or larger end-to-end distance. Importantly, there is a regime (denoted as not feasible) where the QKD cannot be performed since the error rates are large (enough that the secret key fraction (\ref{eq:secret-fraction}) becomes negative) because of the large ratio of the memory idle time over the coherence time. As we see, introducing cutoff would help a bit and shrink the infeasible regime.

\begin{figure}
\centering
    \includegraphics[width=9cm]{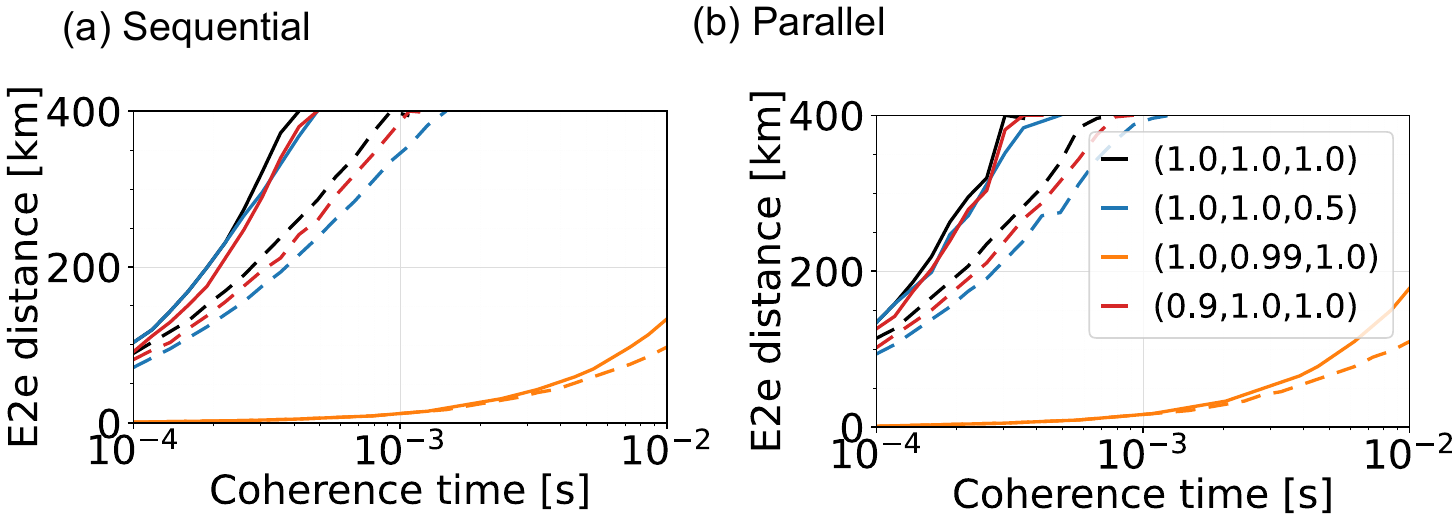}
    \vspace{-0.06in}
  \caption{Feasible regimes of parameters for different values (color coded) of noise parameters $F$ defined in (\ref{eq:link-level-epr}) and $\mu$ in (\ref{eq:2q-depolarizing}) and insertion loss $p_\text{link}$ in (\ref{eq:link-level-prob}). The solid (dashed) lines are the boundaries of the feasible region for the case with (without) cutoff. Each legend indicates the values for $(F, \mu, p_\text{link})$.}
  \label{fig:feasible_region}
\end{figure}

In another experiment here, we compare the boundary of the infeasible regime under different assumptions about the link and node properties. Figure \ref{fig:feasible_region} shows the infeasible regime of our two protocols with and without cutoff  for different values of the noise parameter $\mu$ (associated with link generation and Bell-state measurement, c.f.~(\ref{eq:2q-depolarizing})), the initial fidelity of EPR pairs $F_{i}$, and the insertion loss parameter $p_\text{link}$. Following the same convention as in Figure~\ref{fig:cutoff_coherence_SKR}, we use solid (dashed) lines  to indicate the case with (without) cutoff for the same setup. We observe that the impact of $\mu$ on the infeasible regime is much stronger than the link fidelity or the insertion loss, and that the infeasible region associated with the parallel protocol is smaller. The strong sensitivity to $\mu$ can be understood from the QBER in (\ref{eq:qber}) and the fact that $\mu_{e2e}=\mu^{15}$ according to (\ref{eq:mu-e2e}) which decays very quickly. In contrast, $F$ as it appears in $f_{e2e}$ (\ref{eq:small-f-e2e}) only affects one of QBERs $e_x$ in (\ref{eq:qber}) and is further multiplied by an exponentially small number coming from the decoherence factor in (\ref{eq:small-f-e2e}); as a result, we expect a weaker dependence on $F$ compared to $\mu$.

\subsection{The effect of classical communication delay}

In this experiment, we demonstrate how considering classical communication delay can reduce the overall SKR between the two end nodes. We examine a fixed number of repeaters while increasing the end-to-end distance ($L_{e2e}$) and plot the SKR with and without classical communication delay. In the scenario without classical communication delay, we neglect the time needed to acknowledge entanglement generation and to send the swap results to the end nodes and assume that this information can be exchanged instantaneously.

\begin{figure}
\centering
    \includegraphics[width=8.5cm]{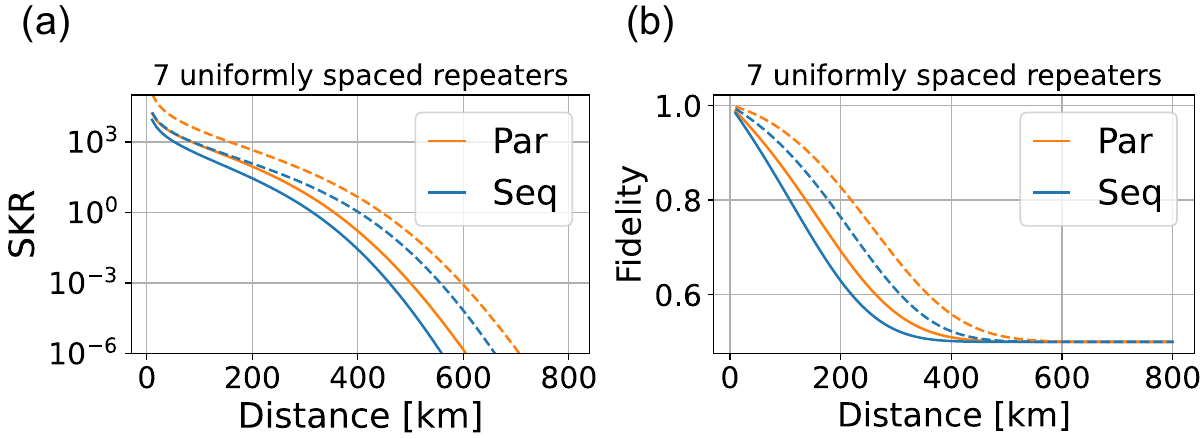}
    \vspace{-0.06in}
  \caption{SKR (a) and fidelity (b) with and without classical comm delay. The solid and dashed lines of the same color represent the same protocol with and without classical comm delay, respectively.}\label{fig:classical_communication_delay}
\end{figure}

Figure \ref{fig:classical_communication_delay} shows the impact of classical communication and how neglecting it gives overestimates for the SKR and fidelity. Here, the solid and dashed lines of the same color represent the same protocol with and without classical communication delay, respectively. As we go to larger distances, the reduction due to the classical communication delay increases further. For example, at $500$km neglecting classical comm causes a significant overestimation where the average performance of each protocol is off by two orders of magnitude. Similarly, the overestimated values of fidelity leads to false impression that entanglement links maintain their quantum properties (i.e.,  $F_{e2e}> 0.5$) up to distances of $400$ km or $600$ km for the sequential and parallel schemes, respectively, while considering the classical communication delays lower these bounds by at least $100$ km.

\subsection{Random repeater placement}

In this experiment, we place $5$ repeaters randomly on a link of length $200$km and plot the distribution of ebit rate and SKR for the two protocols. Since the repeaters are placed randomly, we have a non-uniform repeater chain. The minimum distance between repeaters is required to be $5$ km. 

\begin{figure}
\centering
    \includegraphics[width=8.3cm]{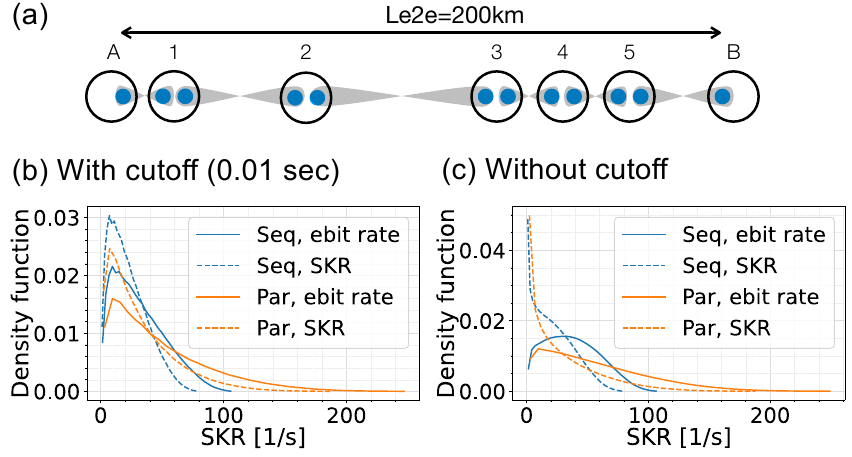}
    \vspace{-0.06in}
  \caption{The distribution of SKR for random placement of $5$ repeaters with and without cutoff. Here, we set $L_{e2e}=200~$km and  $\tau_\text{coh}= 0.1$~sec.}\label{fig:random_repeater_placement_histogram}
\end{figure}

The rationale behind this experiment is to mimic various paths on a large-scale network and compare how different the performance of the two protocols are statistically.
Figure~\ref{fig:random_repeater_placement_histogram}(b) and (c)  show the normalized distribution of ebit and secret key rates for the parallel and sequential protocols with and without a cutoff. As we can see the ebit rates and SKRs follow a similar distribution in each case and largely overlap. Unlike the sequential protocol, the distribution of the rates for the parallel protocol has these long tails towards large values. In other words, the median of the rates for the two protocols  may not differ much although there could be a larger gap in the mean rates. This observation overall implies that the parallel protocol is only marginally better than the sequential one.


\subsection{Different protocols performance in a real network}
In this experiment, we assess the proposed protocols for generating secret keys in an actual network. We select SURFnet and randomly choose user pairs based on their shortest path range. Specifically, we focus on user pairs whose shortest path lengths fall within the range of $[50,350]$ kilometers and include at least $2$ nodes (repeaters). We examine three distinct setups for the sequential and parallel protocols. For each user pair, we seek the optimal cutoff under both sequential and parallel schemes to maximize the secret key rate at various values of the memory coherence time. For each protocol, once the optimal cutoff value for each user pair and given coherence time is determined, we compute the average cutoff value across all user pairs. Subsequently, we evaluate the secret key rate by employing the average cutoff for all user pairs. We also calculate the secret key rate for the sequential and parallel schemes without any cutoff for quantum memories.

\begin{figure}
\centering
    \includegraphics[width=8.8cm]{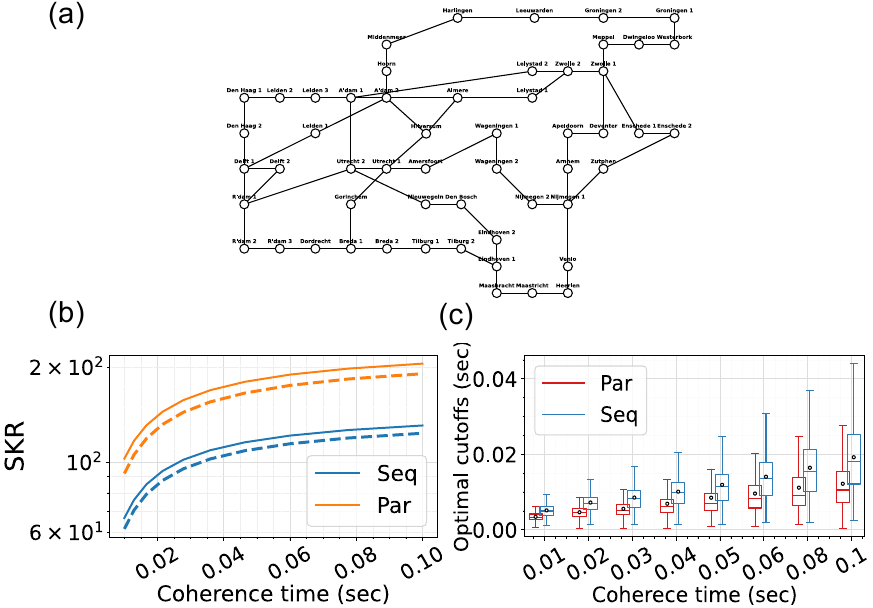}
    \vspace{-0.06in}
  \caption{Secret key rate (b) and optimal cutoff value for quantum memories for $900$ randomly chosen user pairs in SURFnet (c).}
  \label{fig:SURFnet_SKR}
\end{figure}

Figure~\ref{fig:SURFnet_SKR}(a) illustrates the schematic of the SURFnet topology \cite{knight2011internet}, while Figure~\ref{fig:SURFnet_SKR}(b) depicts the SKR as a function of coherence time with and without cutoff values for 900 randomly selected user pairs within SURFnet. An interesting observation is the minimal difference in the SKRs with and without imposing a cutoff on quantum memories irrespective of the protocol type. This suggests that the utilization of distinct cutoff values for different user pairs may not be necessary. In Figure~\ref{fig:SURFnet_SKR}(c), we show a boxplot illustrating the optimal cutoff values for optimizing SKR across 900 user pairs for both protocols. While larger coherence times result in elevated SKRs, user pairs have a more diverse range of optimal cutoff values compared to the case that coherence times are small. This holds significance when dealing with multiple user pairs in the network and deploying repeaters to serve them within a general topology. Such observations, along with others in this paper, should be accounted for during the network planning phase of quantum networks \cite{rabbie2022designing,pouryousef2023quantum}.

\section{Related work}
\label{sec:related work}

Several studies have investigated entanglement generation protocols in the context of repeater chains, with a particular focus on sequential and parallel approaches. Here, we review relevant literature that contributes to the understanding of sequential and parallel entanglement generation in repeater chains.

Kamin \emph{et al.}~\cite{kamin2023exact} analyze the ebit and secret key rates for sequential and another protocol called the “doubling" scheme where the segments are divided into two equal halves and when both halves have finally distributed an entangled state the last swapping is attempted \cite{shchukin2022optimal}. A few papers use Markov Chain and reinforcement learning to obtain the optimal policy for elementary link-level generation \cite{khatri2021policies} or to model the repeater chain evolving states and derive optimal policies for entanglement distribution in repeater chains \cite{haldar2023fast,inesta2023optimal}. A policy will indicate when certain actions such as swap operations should be performed at intermediate repeaters. Goodenough \emph{et al.}~\cite{goodenough2024noise} analyze the expected fidelity with or without cut-off strategies in homogeneous swap as soon as possible (ASAP) repeater chains.  

The majority of the existing research on determining the optimal protocols for entanglement generation in repeater chains relies on either a homogeneous repeater chain, a time-slotted system, or a centralized controller that acquires comprehensive knowledge of the state of all repeaters during each time slot to make decisions regarding the subsequent actions on each repeater. However, obtaining global knowledge can introduce additional classical communication delays to the system and contribute to increased qubit decoherence. Furthermore, these studies do not provide an analysis for scenarios involving a cutoff in inhomogeneous repeater chains and often overlook the impact of classical communication delays in their evaluations.  
        
QCAST \cite{shi2020concurrent} functions on a time-slot basis, utilizing a sequential swapping protocol, and calculates the rate of entanglement generation along a path with more than one memory. Pouryousef \emph{et al.}~\cite{pouryousef2023quantum} embrace a similar sequential swapping approach, but within the context of repeater and memory allocation for a designated set of user pairs during the quantum network planning phase. None of these studies conduct a comprehensive analysis of sequential and parallel protocols or examine the impact of decoherence and cutoff on the rate and SKR. Yang \emph{et al.}~\cite{yang2024asynchronous} introduce a set of asynchronous routing protocols for quantum networks, fostering communication between nodes to augment the entanglement generation rate. However, the extent to which the classical communication overhead of these proposed protocols may influence the rate and SKR remains uncertain.

\section{Conclusion}
\label{sec:conclusion}

In conclusion, our study delves into two asynchronous protocols—sequential and parallel—for entanglement distribution across quantum networks. We introduce two simulation methods to analyze these protocols across various scenarios. The Monte Carlo simulator, though significantly faster, relies on analytical expressions that may become challenging to derive if we add additional complexity to our protocols. Conversely, the discrete event simulator, while slower, closely mirrors actual hardware and readily accommodates protocol modifications.

Through numerous numerical experiments, we find that the parallel protocol generally outperforms the sequential counterpart, albeit with marginal differences considering the potential implementation complexities of the parallel approach. Moreover, the sequential protocol's adaptability to non-deterministic entanglement swaps, encompassing a broader range of quantum memory technologies, and its compatibility with advanced protocols utilizing memory multiplexing, underscore its versatility and potential for enhanced performance.


Our work opens several avenues for future research. 
Firstly, for simplicity in most of our simulations we set some of the noise parameters  to zero in link-level entanglement fidelity, generation and measurement noise models. It would be interesting to study the dependence on these noise parameters in greater details. 
Secondly, incorporating entanglement purification, along with the required classical communication, into our protocols and comparing their performance offers a valuable extension. Lastly, considering the potential of the sequential scheme as a hop-by-hop protocol, an interesting direction is to develop protocols to address routing and congestion control \cite{yang2024asynchronous,xiao2023connectionless,chen2023q,wang2022pre}.








\section{Acknowledgements}
The authors acknowledge insightful discussions with Alireza Shabani, Ramana Kompella, and Raj Jain.
\appendices

\section{Derivation details for sequential protocol}
\label{sec:appendix-sequential} 

In this appendix, we provide some derivation details for the sequential protocol. We consider a linear network with $n$ repeaters placed between two end users and $n+1$ links connecting them. 
We start by noting the moment when each repeater memory is initialized
\begin{align}
    t_{\text{init},k,L} &= \sum_{i=1}^k (2N_i-1)\tau_i, \\
    t_{\text{init},k,R} &= t_{\text{init},k,L}+ 2(N_{k+1}-1)\tau_{k+1}.
\end{align}
Similarly, for the end user memories we have
\begin{align}
    t_{\text{init},A} &= 2(N_1-1)\tau_1, \\
    t_{\text{init},B} &= \sum_{i=1}^{n+1} (2N_i-1)\tau_i.
    \end{align}
As a result, the total time for the end-to-end entanglement generation is given by
\begin{align}
    \label{app:seq-total-time}
    T = t_{\text{init},B} + \tau_{e2e} = \sum_{i=1}^{n+1} 2  \tau_i N_i,
\end{align}
where we added $\tau_{e2e}=\sum_{i=1}^{n+1} \tau_i$ to account for the classical signal transmission (acknowledgement) from $B$ to $A$.

To compute the average fidelity or QBER, we need to find memory idle times individually,
\begin{align}
    T_A &=  T - 2(N_1-1)\tau_1, \qquad  T_B = \tau_{e2e},
\end{align}
for the end memories, and
\begin{align}
    t_{i,L} &= 2 N_{i+1} \tau_{i+1}, \qquad   
    t_{i,R} = 2 \tau_{i+1},
\end{align}
for the repeater memories.
To calculate the expectation value of $f_{e2e}$ in (\ref{eq:small-f-e2e}) we need to evaluate the expectation value of the exponential term $\boldsymbol{E}(e^{-\tau_\text{idle}/\tau_\text{coh}})$.
For the end-to-end fidelity we include the end memory idle times and  the exponent is simplified into
\begin{align}
   \tau_\text{idle}^{(F)} &= 
   T_A + T_B + \sum_{i=1}^{n} (t_{i,L}+t_{i,R}) \nonumber \\
   &= 3 \tau_{e2e} + 4 \sum_{i=2}^{n+1} N_{i} \tau_{i},
   \label{eq:idle-seq-F}
\end{align}
whereas for the SKR we do not include the end memory idle times, and  we obtain
\begin{align}
   \tau_\text{idle}^{\text{(SKR)}} =
      \sum_{i=1}^{n} (t_{i,L}+t_{i,R}) 
    &= 2 \sum_{i=2}^{n+1} (N_{i}+1) \tau_{i}.
    \label{eq:idle-seq-SKR}
\end{align}

In what follows, we discuss the average values of various quantities for both cases with and without the cutoff.

\subsection{No cutoff}

In this case, $N_i$ varies over all positive integers. Hence, the average time for end-to-end entanglement is simply given by
\begin{align}
    \overline{T} = \boldsymbol{E}(T) 
    &= \sum_{i=1}^{n+1} 2\tau_i \cdot \sum_{N_i=1}^\infty N_i p_i (1-p_i)^{N_i-1 } = \sum_{i=1}^{n+1} \frac{2\tau_i}{p_i}.
\end{align}
where in the first equality we use (\ref{app:seq-total-time}). This is a proof for expression (\ref{eq:seq-mean-time}) of the main text.
Next, we calculate the expectation value of the exponential term for the fidelity
\begin{align}
    \label{eq:exp-term-seq-F}
    \boldsymbol{E}(e^{-\tau_\text{idle}^{(F)}/\tau_\text{coh}})
    = e^{-3\tau_{e2e}/\tau_\text{coh}} \prod_{i=2}^{n+1} 
    \left(\frac{p_i e^{-4\tau_i/\tau_\text{coh}} }{1-q_i e^{-4\tau_i/\tau_\text{coh}}}\right),
\end{align}
where $q_i = 1-p_i$. A similar expression can be derived for the exponential term to calculate the SKR,
\begin{align}
    \label{eq:exp-term-seq-SKR}
    \boldsymbol{E}(e^{-\tau_\text{idle}^{(\text{SKR})}/\tau_\text{coh}})
    =  \prod_{i=2}^{n+1}
    \left(\frac{p_i e^{-4\tau_i/\tau_\text{coh}} }{1-q_i e^{-2\tau_i/\tau_\text{coh}}}\right).
\end{align}
We then plug in the above expressions to (\ref{eq:small-f-e2e}) to compute $\overline{f_{e2e}}$ which is ultimately used in (\ref{eq:fidelity}) and (\ref{eq:qber}) to calculate the end-to-end entanglement fidelity and the secret key fraction, respectively.

\subsection{With cutoff}

Unlike the case without any cutoff, here there is a chance that an iteration may not finish with an end-to-end entanglement generation (i.e., no success). We start with the one repeater case as a warm-up example and next present the general case with $n$ repeaters.

\subsubsection{One repeater}

The protocol for two segments (one repeater) is as follows:
\begin{itemize}
    \item[1.] We try the first link and repeat until we succeed. Then we go to step 2.
    \item[2.] We try the second link only $m$ times, where $m = [\frac{\tau_\text{cut}}{2\tau_2}]$, $\tau_\text{cut}$ is the cutoff time, and $[x]$ is the integer part of $x$. If we succeed we claim a successful event, otherwise it is a failure and we go to step 1 and start from the beginning. This is to ensure that the idle time of the left memory inside the repeater does not exceed the cutoff time.
\end{itemize}
To compute the average time, we need to enumerate successful events, how long they take, and what is their success probability.
Hence, the average time is found by
\begin{align}
    \overline T^{\text{(cut)}}_2 &= \sum_{r=1}^\infty P_{2,m} Q_{2,m}^{r-1} \left( \frac{2r\tau_1}{p_1}+  (r-1) \tau_\text{cut}+ \frac{2 N_m(p_2)}{P_{2,m}}\tau_2 \right) \nonumber \\
    &=
    \frac{2\tau_1}{p_1 P_{2,m}}
    +  \left(\frac{1}{P_{2,m}}-1 \right)\tau_\text{cut} + \frac{2N_m(p_2)}{P_{2,m}} \tau_2,
    \label{eq:seq_cutoff}
\end{align}
where $Q_{2,m}= (1-p_2)^m$, $P_{2,m}=1-Q_{2,m}$, and $N_m(\cdot)$ is given by
\begin{align}
    \label{eq:Ntrial_m}
    N_m (p) &= \sum_{N=1}^m p q^{N-1} N 
    = \frac{1-(1+mp)q^m}{p}.
\end{align}
It is easy to see that as $m\to \infty$ we obtain $\overline T^{\text{(cut)}}_2 \to 2(\frac{\tau_1}{p_1}+\frac{\tau_2}{p_2})$.

\subsubsection{$n$ repeaters}
Following the idea behind the underlying ebit generation protocol (that is sequential) we can calculate the average total time progressively. We have seen how to calculate the total time for two segments. Each term in the sum in (\ref{eq:seq_cutoff}) represents a successful event after trying the sequential scheme with cutoff $r$-times. Hence, we calculate the average time of these events by plugging the average total time for the first link (that is $\frac{2\tau_1}{p_1}$) and add the contribution from the second link (which is subject to the cutoff policy). Similarly, we can write a recursion relation for the average time of delivering entanglement to the $k$-th repeater in terms of the average time of delivering entanglement to the $(k-1)$-th repeater as follows
\begin{align}
    T^{\text{(cut)}}_{k} = \frac{T^{\text{(cut)}}_{k-1}}{P_{k,m_k}} +  \left(\frac{1}{P_{k,m_k}}-1 \right)\tau_\text{cut} + \frac{2N_{m_k}(p_k)}{P_{k,m_k}} \tau_k,
\end{align}
where $Q_{k,m_k}= (1-p_k)^{m_k}$, $P_{k,m_k}=1-Q_{k,m_k}$, and $m_k=[\frac{\tau_\text{cut}}{2\tau_k}]$ correspond to the $k$-th segment between the $k$-th and $(k-1)$-th repeaters. In our analysis, we numerically calculate the average end-to-end ebit generation time $T^{\text{(cut)}}_n$ recursively by using the above equation starting with $T^{\text{(cut)}}_1 = \frac{2\tau_1}{p_1}$.

Next, we compute the expectation value of the exponential term for the fidelity and SKR. We note that the overall memory idle time still follows (\ref{eq:idle-seq-F}) and (\ref{eq:idle-seq-SKR}), but are subject to the constraints $N_i \leq m_i = [\frac{\tau_\text{cut}}{2\tau_i}]$. Hence, we truncate the sum as in $\boldsymbol{E}^{\text{(cut)}}(x_{N_i}) = P_{i,m_i}^{-1} \sum_{N_i=1}^{m_i} p_i (1-p_i)^{N_i-1} x_{N_i}$ where the prefactor is a normalization constant introduced earlier. As a result, we obtain
\begin{align}
    &\boldsymbol{E}^{\text{(cut)}}(e^{-\tau_\text{idle}^{(F)}/\tau_\text{coh}})  \\
    &= e^{-3\tau_{e2e}/\tau_\text{coh}} \prod_{i=2}^{n+1} 
    \left(\frac{p_i e^{-4\tau_i/\tau_\text{coh}}}{P_{i,m_i}} 
    \frac{ (1-q_i ^{m_i}e^{-4m_i\tau_i/\tau_\text{coh}})}{(1-q_i e^{-4\tau_i/\tau_\text{coh}})}\right),\nonumber
\end{align}
for the fidelity and
\begin{align}
    &\boldsymbol{E}^{\text{(cut)}}(e^{-\tau_\text{idle}^{(\text{SKR})}/\tau_\text{coh}})\\
     &=  \prod_{i=2}^{n+1}
    \left( \frac{p_i e^{-4\tau_i/\tau_\text{coh}} }{P_{i,m_i}}
     \frac{(1-q_i ^{m_i}e^{-2m_i \tau_i/\tau_\text{coh}})}{(1-q_i e^{-2\tau_i/\tau_\text{coh}})}\right).\nonumber
\end{align}
for the SKR. We note that in both cases taking $m_i\to \infty$ we recover (\ref{eq:exp-term-seq-F}) and (\ref{eq:exp-term-seq-SKR}), respectively.

\section{Derivation details for parallel protocol}
\label{sec:appendix-parallel} 

In this appendix, we derive explicit equations for the swapping schedule and memory idle time in terms of set of random variables $N_i$ which denote the number of attempts until we get a successful entanglement link as explained in Sec.~\ref{sec:noise model}. To evaluate the average quantities, we numerically sample $N_i$ according to the geometric distribution $p_i (1-p_i)^{N_i}$. Here, we start with the case of one repeater as a warm up example and present the general case afterward.

\subsubsection{one repeater}

The entanglement swapping is performed at $ T_\text{sw}= \max( (2N_1-1) \tau_1, 2 N_2 \tau_2)$. The total time is then given by
$T= \tau_1 + T_\text{sw}$,
where $\tau_1$ is added to account for the classical communication to send the entanglement swapping measurement outcome to the sender.
The idle time for the end-node memories are found to be
\begin{align}
    T_A &= T - 2 (N_1-1) \tau_1, \\
    T_B &= T - (2 N_2-1) \tau_2,
\end{align}
and for network memories are given by
\begin{align}
    t_{1,L} &= T_\text{sw} - (2N_1-1) \tau_1, \\
    t_{1,R} &= T_\text{sw} - 2(N_2-1) \tau_2.
\end{align}
Hence, the total memory idle time is 
\begin{align}
    T_A+ T_B +t_{1,L}+ t_{1,R} =
    4T - (4N_1-1)\tau_1 - (4N_2-3)\tau_2.
\end{align}

\subsubsection{n repeaters}

In this case, there are $n+1$ elementary links connecting the $n$ repeaters. 
We first note that the entanglement swapping at repeater $k$ occurs at
\begin{align}
    T_k= \max((2N_k-1) \tau_k, 2 N_{k+1} \tau_{k+1}).
\end{align}
as explained in the main text.
The total time is given by the last classical signal which is the measurement outcome of the entanglement swapping arriving at the sender as in (\ref{eq:T-parallel}).
Furthermore, the idle time of memories in the $k$-th repeater between segments $k$ and $k+1$
is given by
\begin{align} 
\label{eq:tkl-par}
    t_{k,L} &= T_k - (2N_k-1) \tau_k, \\
\label{eq:tkr-par}
    t_{k,R} &= T_k - 2(N_{k+1}-1) \tau_{k+1}.
\end{align}
Hence, we have
\begin{align}
    t_{k,L}+t_{k,R} &= 2T_k - (2N_k-1) \tau_k - 2(N_{k+1}-1) \tau_{k+1} \nonumber \\
    &= |(2N_k-1) \tau_k- 2 N_{k+1} \tau_{k+1})|+ 2\tau_{k+1}.
\end{align}
The idle time for the end-node memory are given by
\begin{align}
    T_A &= T - 2 (N_1 -1 )\tau_1, \\
    T_B &= T - (2N_{n+1} -1 )\tau_{n+1}.
\end{align}
For each random iteration, we plug in the numerical values of the above expressions to (\ref{eq:small-f-e2e}) to compute $\overline{f_{e2e}}$ which is ultimately used in (\ref{eq:fidelity}) and (\ref{eq:qber}). Next, we average over them to numerically calculate the expectation values of the end-to-end entanglement fidelity and the secret key fraction, respectively.

We introduce the cutoff by imposing constraints $t_{k,s}\leq \tau_\text{cut}$ with $s=L,R$ on the memory idle times (\ref{eq:tkl-par}) and (\ref{eq:tkr-par}) and discard those attempts which violate the constraints.

\bibliographystyle{IEEEtran}
\bibliography{refs.bib} 

\end{document}